\documentstyle[12pt]{article}
\begin{document}

\pagestyle{plain}

\vspace*{-10mm}

\baselineskip18pt
\begin{flushright}
{\bf BROWN-HET-940}\\
{\bf hep-ph/9403406}\\
{\bf March 1994}\\
\end{flushright}
\vspace{1.0cm}
\begin{center}
{\Large \bf Electroweak Radiative Corrections,}\\
{\Large \bf Born Approximation, and Precision Tests}\\
{\Large \bf of the Standard Model at LEP}\\
\vglue 5mm
{\bf Kyungsik Kang } \\
\vglue 2mm
{\it Department of Physics, Brown University,
Providence, RI 02912, USA,\footnote{Permanent address and supported
in part by the USDOE contract DE-FG02-91ER40688-Task A} }\\
\vglue 1mm
{\it and}\\
\vglue 1mm
{\it Division de Physique Th\'eorique, Institut de Physique Nucl\'eaire,
91406 Orsay Cedex and }\\
\vglue 1mm
{\it LPTPE, Universit\'e P. \& M. Curie, 4 Place Jussieu, 75252 Paris
Cedex05, France }\\
\vglue 3mm
{\it and}\\
\vglue 3mm
{\bf Sin Kyu Kang } \\
\vglue 2mm
{\it Center for High Energy Physics, Korea Advanced Institute of }\\
\vglue 1mm
{\it Science and Technology, Taejon, Korea }\\
\vglue 10mm
{\bf ABSTRACT} \\
\vglue 8mm
\begin{minipage}{14cm}
{\normalsize
We have examined the evidence for the electroweak radiative
corrections in the LEP precision data along with the intriguing
possibility that the QED corrections only may be sufficient
to fit the data.
We find that the situation is very sensitive to the precise value
of $M_W$.
While the world average value of $M_W$ strongly favors nonvanishing
electroweak radiative corrections, the QED corrections alone
can account for the data within $2\sigma $ in the context of the
standard model.
We discuss how the precision measurements of $M_W$ can provide a decisive
test for the standard model with radiative corrections
and give a profound implication for the existence of t-quark and Higgs
scalar.
}
\end{minipage}
\end{center}
\newpage
%
%
Recently much interests have been paid to the electroweak radiative
corrections (EWRC)
and precision tests of the standard model thanks to the accurate data
obtained at LEP [1-7 ].
       There have been
       numerous articles published on the subject as has been documented
in [7,8].
The LEP data are generally   regarded as the success of the standard model
and as the  evidence  for the nonvanishing EWRC [9 ].
However, Novikov, Okun, and Vysotsky [10 ] have argued recently that the
experimental
data from LEP on the electroweak parameters as defined in the standard model
could be explained by the Born approximation with     $\alpha (M^2_Z)$
instead of $\alpha(0)$ and the corresponding redefinition of the weak
mixing angle $\sin ^2\theta $ instead of $\sin ^2\theta_W $ and that
the genuine EWRC are yet
to be observed.
In particular the electroweak Born predictions
are claimed to be within $1\sigma $ accuracy of all electroweak precision
measurements made at LEP.
This is very interesting because    no Born approximation in any precision
test has ever produced such an impressive description of all  available data
in our memory.

In this paper, we reexamine this claim and test if the present LEP data
[1-5]  can indeed be accounted for by the QED corrections only of the full
one-loop EWRC, i.e., by the electroweak Born approximation (EWBA).
In order to do so, we have firstly   considered
the case    of the pure QED corrections by consistently turning off the
non-photonic one-loop
contributions coming from the weak interaction origin
in the full one-loop EWRC and
then compared the results with the full one-loop EWRC
with the aid of the ZFITTER program [11] but with a few
modifications such as using an improved QCD correction factor and
making the best $\chi^2$ fit to the data.

Since the basic lagrangian contains the bare electric charge $e_0$,
the renormalized physical charge $e$ is fixed by a counter term $\delta e$;
$  e_0 = e + \delta e$.
The counter term $\delta e$ is determined by the condition of the on-shell
charge renormalization in the $\overline{MS}$ or on-shell scheme.
It is well known that the charge renormalization in the conventional QED
fixes the counter term by
the renormalized vacuum polarization $\hat {\Pi}^{\gamma }(0)$ and
one can evaluate
$\hat {\Pi}^{\gamma}(q^2)=\hat {\Sigma} ^{\gamma \gamma }(q^2)/q^2$
from the photon self energy $\hat{\Sigma}^{\gamma \gamma }(q^2)$, for example,
by the dimensional regularization method. This gives at $q^2=M_Z^2$
the total fermionic contribution of $m_f \leq M_Z $ to the real part
$ Re\hat {\Pi}^{\gamma}(M_Z^2) = -0.0602(9)$,
so that the {\it running} charge defined as
\begin{equation}
e^2(q^2) = \frac{e^2}{1+Re\hat {\Pi }^{\gamma}(q^2)}
\end{equation}
gives $\alpha(M_Z^2)=1/128.786 $ in the on-shell scheme if the hyperfine
structure constant
$\alpha = e^2/4\pi = 1/137.0359895(61)$
is used.
The concept of the running charge, however, is scheme dependent [12]
: the
$\overline {MS}$ fine structure constant at the $Z$-mass scale is given by
\begin{equation}
   \hat{\alpha }(M_Z)=\alpha /[1-\Pi^{\gamma}(0)|_{\overline {MS}}+
  2\tan {\theta_W}(\Sigma^{\gamma Z}(0)/M^2_Z)_{\overline{MS}}].
\end{equation}
In this case, one can show
                $\hat{\alpha}(M_Z) = (127.9\pm 0.1)^{-1}$,
which is different by some 0.8 $\%$ from the on-shell $\alpha(M_Z^2)$,
as $\hat{\alpha}(M_Z)$ gets the weak gauge boson contributions also.

Taking just the QED one-loop contributions of the photon self-energy in
the full EWRC is equivalent to the EWBA with the effective $\alpha
(M^2_Z)$ instead of $\alpha $ in the sense of Novikov et al. But unlike
our QED case, they substituted $\alpha (M^2_Z)$ also for $\alpha $ in the
Coulomb correction factor $R_{QED}$ originating from the real photon
emissions from the external states.

The electroweak parameters are evaluated
numerically with the hyperfine structure constant $\alpha$,
the four-fermion coupling constant of $\mu$-decay,
$G_{\mu} = 1.16639(2)\times 10^{-5} \mbox{GeV}^{-2}$, and $Z$-mass,
$M_Z = 91.187(7) \mbox{GeV}$.
Numerical estimate of the full EWRC requires
the mass values of the leptons, quarks, Higgs scalar and $W$-boson besides
 these
quantities.
While $Z$-mass is known to an incredible accuracy from the LEP experiments
largely due to the resonant depolarization method, the situation with the
$W$-mass is desired to be improved, i.e., $M_W = 80.22(26)~$ GeV [13] vs.
the     CDF measurement $M_W = 79.91(39)~$ GeV [14] and
$M_W/M_Z = 0.8813(41) $ as determined by UA2 [15].
One has, in the standard model, the on-shell
relation
\begin{equation}
\sin^2 \theta_W = 1-\frac{M_W^2}{M_Z^2},
\end{equation}
while the four-fermion coupling constant $G_{\mu}$
can be written as
\begin{equation}
G_{\mu } = \frac{\pi \alpha }{\sqrt{2}M_W^2}
\left(1-\frac{M_W^2}{M_Z^2}\right)^{-1}(1-\Delta r)^{-1}
\end{equation}
so that $\Delta r$, representing the radiative corrections, is given by
\begin{equation}
\Delta r = 1-\left(\frac{37.28}{M_W}\right)^2
\frac{1}{1-M_W^2/M_Z^2}.
\end{equation}
We note from Table 1(Cases 1,2) that the radiative correction $\Delta r$
is very sensitive
to the value of $M_W$.
Mere change in $M_W$ by $0.39\%$ results as much as a $28\%$ change in
$\Delta r$.
\begin{table}
\begin{center}
\begin{tabular}{|c|c|c|c|}\hline \hline
 & $M_W$ (GeV) & $\Delta r$ & $\sin^2 \theta_W$ \\ \hline
 1& 79.91 & 0.0623 & 0.2321 \\ \hline
 2& 80.22 & 0.0448 & 0.2261 \\ \hline \hline
 &  $\sin^2 \theta_W$ & $\Delta r $ & $M_W$ (GeV) \\ \hline
 3 & 0.2257 & 0.0436 & 80.239 \\ \hline
 4 & 0.2319 & 0.0617 & 79.917 \\
\hline \hline
\end{tabular}
\caption{ Dependence of the radiative correction $\Delta r $
on the values of $M_W$ (Cases 1,2) and of $\sin^2 \theta_W$ (Cases 3,4).}
\end{center}
\end{table}

On the other hand, the QED contribution to $\Delta r$ is
$(\Delta r )_{\mbox{QED}}
= -Re \hat{\Pi }^{\gamma}(M_Z^2)=0.0602.$
Thus we see from Table 1(Cases 1,2) that the QED portion of $\Delta r$ is
a major component of the radiative corrections, particularly in the case
of CDF $M_W$, for which the QED contribution is already within $3.34\%$ of
the needed
$\Delta r$ and is close enough to be within the
experimental uncertainty.
However, with the current world average value $M_W=80.22 $ GeV, the QED
corrections leave $34.4\%$ that is
to be accounted for by the weak interaction corrections.

Using $M_Z$ and $\sin^2 \theta_W$ instead of $M_W$, $\Delta r$ can also be
expressed as
\begin{equation}
\Delta r = 1-\frac{(\frac{\pi \alpha}{\sqrt{2}G_{\mu}})}{M_Z^2
             \cos^2 \theta_W\sin^2 \theta_W}
         = 1-\frac{0.16714}{\cos^2 \theta_W\sin^2 \theta_W}.
\end{equation}
Table 1(Cases 3,4) shows the estimates of $\Delta r$ as well as $M_W$ for
two values of the on-shell
weak-mixing angle, i.e., $\sin^2 \theta_W = 0.2319$ [16] based on quark
charge asymmetry
or forward-backward asymmetry measurements at LEP and
$\sin^2\theta_W=0.2257$ [5] as summarized by LEP collaborations.
We see
that precise determination of the on-shell value of $\sin ^2\theta_W$ can
also constrain the needed radiative correction and the value of $M_W$,
thus providing another crucial test for the evidence of the EWRC in the
standard model.

We have examined the full EWRC to the nine
observables of the Z-decay and $M_W$ as shown in Table 2 and 3. Full
details with complete results of theoretical formula of renormalization
and full EWRC will be presented elsewhere [17]. These parameters are
calculated with a modified ZFITTER program in which the best $\chi^2$
fit to the data is
searched and the gluonic coupling constant
$\bar{\alpha _s}(M_Z^2) = 0.123 \pm 0.006$
is used in the improved QCD correction factor [18]
$R_{\mbox{QCD}} = 1+1.05\frac{\bar{\alpha_s}}{\pi}+0.9(\pm 0.1)
\left(\frac{\bar{\alpha_s}}{\pi}\right)^2-13.0
\left(\frac{\bar{\alpha_s}}{\pi}\right)^3$
for all quarks that can be produced in the $Z\rightarrow
f\bar{f}$ decay. The partial width for $Z\rightarrow f\bar{f}$ is given by
\begin{equation}
\Gamma_f = \frac{G_{\mu}}{\sqrt{2}}\frac{M_Z^3}{24\pi}\beta R_{\mbox{QED}}
c_fR_{\mbox{QCD}}(M_Z^2)\left \{ [(\bar{v}^Z_f)^2+(\bar{a}^Z_f)^2]\times
 \left(1+2\frac{m_f^2}{M_Z^2}\right)-6(\bar{a}^Z_f)^2\frac{m_f^2}
{M_Z^2}\right \}
\end{equation}
where $\beta =\beta(s)=\sqrt{1-4m_f^2/s}$ at $s=M_Z^2$, $R_{\mbox{QED}}
=1+\frac{3}{4}\frac{\alpha}{\pi}Q_f^2$ and the color factor $c_f=3$ for
quarks and 1 for leptons.
Here the renormalized vector and axial-vector couplings are defined by
$\bar{a}_f^Z=\sqrt{\rho_f^Z}2a_f^Z = \sqrt{\rho_f^Z}2I_3^f $ and
$\bar{v}^Z_f=\bar{a}^Z_f[1-4|Q_f|\sin^2\theta_W\kappa^Z_f] $ in
terms of the familiar notations [11, 12].
Note that the QED correction $(\Delta r)_{QED}$  is included in the
couplings through $\sin^2 \theta_W$ via (3) and (5) and all other
non-photonic loop corrections are
grouped in $\rho_f^Z$ and $\kappa_f^Z$ as in [11,17,19].
Thus the case of the QED corrections only, i.e., the EWBA
can be achieved simply by setting $\rho^Z_f$ and $\kappa^Z_f$
to 1 in the vector and axial-vector couplings.
\begin{table}
\begin{center}
\begin{tabular}{|c||c|c||c|c|c|} \hline \hline
 & Experiment & Pure QED correc. & Full EW & Full EW & Full EW \\
  \hline
 $m_t$~(GeV) & 150  & & 120 & 138 & 158 \\
 $ m_H$~(GeV) & 60 $\leq m_H \leq 1000$ & & 60 & 300 & 1000 \\
 \hline
$M_W$~(GeV) & $79.91\pm 0.39$ & 79.94 & 80.10 & 80.10 & 80.13 \\
$ \Gamma_Z $~(MeV) & $2488.0\pm 7.0 $ & 2488.4 & 2489.0 & 2488.9 & 2488.8 \\
$ \Gamma_{b\bar{b}} $~(MeV) & $383.0\pm 6.0 $ & 379.4 & 377.4 & 376.5 & 375.4
\\
$ \Gamma_{l\bar{l}} $~(MeV) & $83.52\pm 0.28 $ & 83.47 & 83.53 & 83.53
& 83.63 \\
$ \Gamma_{had} $~(MeV) & $1739.9\pm 6.3 $ & 1740.3 & 1738.8 & 1738.2 & 1737.7
\\
$R(\Gamma_{b\bar{b}}/\Gamma_{had})$ & $0.220 \pm 0.003$ & 0.218 & 0.217 &
0.217 & 0.216 \\
$R(\Gamma_{had}/\Gamma_{l\bar{l}})$ & $20.83 \pm 0.06$ & 20.85 & 20.82 &
20.81 & 20.78 \\
$\sigma _h^P (nb)$ & $41.45 \pm 0.17 $ & 41.41 & 41.37 & 41.38 & 41.40 \\
$ g_V $ & $-0.0372 \pm 0.0024 $ & -0.0372 & -0.0341 & -0.0334 & -0.0334 \\
$ g_A $ & $-0.4999 \pm 0.0009 $ & -0.5000 & -0.5003 & -0.5005 & -0.5006 \\
\hline
$ \sin ^2{\theta_W}$ & 0.2321 & 0.2314 & 0.2284 & 0.2283 & 0.2278 \\
\hline
$ \Delta r $ & 0.0623 & 0.06022 & 0.05162 & 0.05131 & 0.04967 \\
\hline \hline
\end{tabular}
\caption{Numerical results including full EWRC for
nine experimental parameters of the Z-decay and $M_W$. The case of pure
QED corrections only, i.e., EWBA is shown also for comparison.
Each pair of $m_t$ and $m_H$ represents the case of the best $\chi ^2$
for the given input $m_H$ and experimental $M_W = 79.91\pm 0.39 $ GeV.}
\end{center}
\end{table}
\begin{table}
\begin{center}
\begin{tabular}{|c||c|c||c|c|c|} \hline \hline
 & Experiment & Pure QED correc. & Full EW & Full EW & Full EW \\
\hline
 $m_t$~(GeV) & 150  & & 126 & 142 & 160 \\
 $ m_H$~(GeV) & 60 $\leq m_H \leq 1000$ & & 60 & 300 & 1000 \\
 \hline
$M_W$~(GeV) & $80.22\pm 0.26$ & 79.94 & 80.13 & 80.13 & 80.15 \\
$ \Gamma_Z $~(MeV) & $2488.0\pm 7.0 $ & 2488.4 & 2490.2 & 2489.7 & 2489.3 \\
$ \Gamma_{b\bar{b}} $~(MeV) & $383.0\pm 6.0 $ & 379.4 & 377.3 & 376.4 & 375.3
\\
$ \Gamma_{l\bar{l}} $~(MeV) & $83.52\pm 0.28 $ & 83.47 & 83.53 & 83.63
& 83.63 \\
$ \Gamma_{had} $~(MeV) & $1739.9\pm 6.3 $ & 1740.3 & 1739.6 & 1738.8 & 1738.1
\\
$R(\Gamma_{b\bar{b}}/\Gamma_{had})$ & $0.220 \pm 0.003$ & 0.218 & 0.217 &
0.216 & 0.216 \\
$R(\Gamma_{had}/\Gamma_{l\bar{l}})$ & $20.83 \pm 0.06$ & 20.85 & 20.83 &
20.79 & 20.78 \\
$\sigma _h^P (nb)$ & $41.45 \pm 0.17 $ & 41.41 & 41.38 & 41.39 & 41.40 \\
$ g_V $ & $-0.0372 \pm 0.0024 $ & -0.0372 & -0.0344 & -0.0337 & -0.0335 \\
$ g_A $ & $-0.4999 \pm 0.0009 $ & -0.5000 & -0.5004 & -0.5006 & -0.5007 \\
\hline
$\sin^2 {\theta_W}$ & 0.2261 & 0.2314 & 0.2278 & 0.2279 & 0.2275 \\
\hline
$ \Delta r $ & 0.0448 & 0.06022 & 0.04975 & 0.04991 & 0.04895 \\
\hline \hline
\end{tabular}
\caption{The same as Table 2 but for the experimental
$M_W = 80.22 \pm 0.26$ GeV.}
\end{center}
\end{table}
Numerical results for the best $\chi^2$ fit to the experimental
parameters of $Z$-decay
are shown in Tables 2 and 3
for
$M_W = 79.91\pm 0.39 $ GeV and $ M_W = 80.22 \pm 0.26$ GeV respectively
as experimental inputs. They correspond to the values that give
the best $\chi^2$ fit in each case.
Also included in the Tables 2 and 3 is the case of pure QED corrections,
i.e., the EWBA, as well as the output $\sin^2 \theta_W$ and $\Delta r$,
for comparison.
We see that the contributions of the weak corrections are
generally small and the QED portion of the radiative corrections
seems to be close to the experimental values
within the uncertainty of the current measurements. The smallness of the weak
corrections is achieved by a precarious compensation of two
large contributions i.e., between those of t-quark and Higgs scalar.
In general, the radiative correction parameter $\Delta r$ can be written as
\begin{equation}
\Delta r = (\Delta r)_{\mbox{QED}}-\frac{c^2_W}{s_W^2}\Delta \rho
+\Delta r_{rem}
\end{equation}
where $c^2_W$ and $s_W^2$ denote $\cos^2 \theta_W$ and $\sin^2 \theta_W$.
Main contribution to $\Delta \rho =\rho -1$ is from the heavy t-quark
through the mass renormalizations of weak gauge bosons $W $ and $Z,$
while there is a part in $(\Delta r)_{rem}$ containing also the t-quark
and Higgs scalar contributions.
The near absence of the weak interaction contributions to the
radiative corrections is more impressive for $M_W$ = 79.91 GeV than for $M_W$ =
80.22 GeV. At closer examination, however, the EWBA
in the latter case over-estimates the radiative corrections and the full
one-loop EWRC fair better.

In order to complete the analysis of the global fit to the data,
$m_H$ is allowed to vary in the range $60-1000 $ GeV. We find the best fit to
the data is obtained by $m_t = 142^{-16}_{+18} $ GeV, the central value
being the best $\chi^2$ case of $m_H=300$ GeV in Table 3. The
upper and lower bounds on $m_t$ as a function of $m_H$ are shown in Fig. 1
and Fig. 2 for the two experimental $M_W$ at several confidence levels(CL).
For example, in the case of experimental $M_W$ = 80.22 GeV,
the upper bound on $m_t$ is 198 (181) GeV at $95\%$ ($68 \%$) CL, while the
lower bound on $m_t$ is $112 (135)$ GeV [20] at $95 (68)\%$ CL that come
from $m_H=1000 $ GeV.

Note from Tables 2 and 3 that the best global fits to the data give a rather
stable output $M_W=80.13\pm 0.03~$ GeV if the full EWRC are
taken into account, which
is to be contrasted to the output $M_W=79.94$ GeV for the EWBA,
for either experimental $M_W$ value. Also
$\sin^2 \theta_W=0.2279\pm 0.0005$ in the case of the full EWRC is
to be compared to $\sin ^2\theta _W=0.2314$
in the case of QED corrections only.
Clearly one needs better precisions on the measurements of $M_W$.
While the current world average value of
$M_W$ supports strongly for the evidence of the full EWRC
in the LEP data, the Born approximation appears to be in fair
agreement, i.e., within $2\sigma$, with the data at the present precisions.
In order to dismiss the Born approximation, an improvement
of better than $100$ MeV in the error of $M_W$ over the current data is
desired. If $M_W$ turns out to be definitely at around 79.94 GeV with such
precision, then the QED correction is all that has been observed at LEP and
one is cultivating the null result of the EWRC to produce the
range of t-quark mass as pointed out in [10]. However, if $M_W           $ is
definitely around 80.13 GeV with the desired precision, the existing LEP data
are the evidence for the nonvanishing electroweak radiative corrections.

We have examined the results of the best $\chi^2$ fit to the precision
measurements of the Z-decay parameters at LEP and $M_W$ in the standard
model
with the full EWRC as well as those of the EWBA, i.e., the QED
corrections only with the aid of a modified ZFITTER program.
We find that the Born
approximation is in agreement with the data within $2\sigma$ level of
accuracy at the present state of precision while the world
average value of $M_W$ clearly
supports for the evidence of the nonvanishing EWRC in the LEP data.
Further precision measurement of $M_W$ can provide
a real test of the standard model as it will give a tight constraint for the
needed amount of the EWRC providing a profound implication to the mass of
t-quark and Higgs scalar in the context of the standard model.
As long as t-quark remains unobserved, either amount of the radiative
corrections,i.e., $\Delta r \simeq 0.05$ with the full EWRC
and $\Delta r = \Delta \alpha \simeq 0.06$ with the QED
corrections only, can fit the data more or less equally well. But if $M_W$
is determined within 100 $MeV$ uncertainty, $\Delta r$ within the
context of the standard model will be tightly constrained to distinguish
the evidence for the radiative corrections that can discriminate the
mass range of the t-quark and Higgs scalar, thus
providing a crucial test for and even the need of new physics beyond the
standard model.
\section*{Acknowledgements}
One of us (KK) would like to thank the Center for
Theoretical Physics, Seoul National University (CTPSNU) and
the Korea Advanced Institute of Science and Technology (KAIST) where
parts of the work were done and LPTPE, Universit\'e P.\&M. Curie where
the work was completed for the kind hospitality during his sabbatical
stay. Also the authors would like to thank
Professors Hi-sung Song, Jae Kwan Kim, R. Vinh Mau and other colleagues
at CTPSNU, KAIST, and LPTPE for the stimulating environment and supports
and in particular Professor M. Lacombe
for checking the numerical computations.

\section*{References}
\begin{description}
\item[1.] ALEPH Collab., D.Buskulic et al., CERN-PPE/93-40 (1993).
\item[2.] DELPHI Collab., D.Aarnio te al., Nucl.Phys. {\bf B367}
          (1911) 511.
\item[3.] L3 Collab., O.Adriani et al., CERN-PPE/93-31 (1993).
\item[4.] OPAL Collab., P.D.Action et al., CERN-PPE/93-03 (1993).
\item[5.] The LEP Collab., CERN-PPE/93-157 (1993).
\item[6.] C.DeClercq; V.Innocente; and
          R.Tenechinl, in: Proc. XXVIIIth Recontres de Moriond
          (Les Arcs, 1993).
\item[7.] See also L.Rolandi, in: Proc. XXVI ICHEP 1992, CERN-PPE/92
          -175 (1992); M.P.Altarelli, talk given at Les Rencontres
          de Physique de la Vallee d'Aoste (La Thuil, 1993),
          LNF-93/019(p); and J.Lefranceis, in: Proc.
          Int. Europhys. Conf. H. E. Phys. (Marseille, 1993).
\item[8.] F.Dydak, in: Proc. 25 Int. Conf. H. E. Phys.,
          Eds. K.Phua, Y.Tamaguchi (World Scientific, Singapore, 1991),p3.
\item[9.] W.J.Marciano, Phys. Rev. {\bf D 20} (1979) 274;
          A.Sirlin, Phys.Rev. {\bf D22} (1980) 971; (1984) 89;
          A.Sirlin and W.J.Marciano, Nucl.Phys.{\bf B189} (1981) 442;
          and A.Sirlin, NYU-TH-93/11/01. See also W. Hollik, in:
          {\it Precision Tests of the Standard Model}, ed. P.
          Langacker (World Scientific Pub., 1993).
\item[10.] V.A.Novikov, L.B.Okun and M.I.Vysotsky, Mod.Phys.Lett.
          {\bf A8} (1993) 5929.
\item[11.] D.Bardin et al., CERN-TH-6443-92 (1992).
\item[12.] See, for example, W. Hollik (Ref. 9); and K. Kang, in: Proc.
           14th Int. Workshop Weak Interactions and Neutrinos (Seoul,
           1993), Brown-HET-931 (1993).
\item[13.] Particle Data Group, Review of Particle Properties,
          Phys.Rev.{\bf D 45}, No.11, Part II (1992).
\item[14.] CDF Collab., F.Abe et al., Phys.Rev.{\bf D 43} (1991) 2070.
\item[15.] UA2 Collab., J. Allitti et al., Phys. Lett. {\bf B 276}
           (1992) 354.
\item[16.] M. P. Altarelli (Ref. 7).
\item[17.] Kyungsik Kang and Sin Kyu Kang (to be published)
\item[18.] T.Hebbeker, Aachen preprint PITHA 91-08 (1991); and
           S.G.Gorishny, A.L.Kataev and S.A.Larin, Phys.Lett.{\bf B 259}
           (1991) 144. See also L.R.Surguladze and M.A.Samuel,
           Phys.Rev.Lett.{\bf 66} (1991) 560.
\item[19.] M. Consoli and W. Hollik, in {\it Z Physics at LEP 1}, Vol. 1,
           eds. G. Altarelli et al., CERN 89-08 (1989).
\item[20.] The most recent CDF lower bound on $m_t$ is 112 GeV.
           See A.Barbaro-Galtieri, in: Proc. Int. Europhys.
           Conf. on H. E. Phys. (Marseille, 1993).

\section*{Figure Captions}
\item[Fig. 1]: The mass ranges of $m_t$ and $m_H$ at several Confidence
               Levels for $M_W$ = 79.91 GeV.
\item[Fig. 2]: The same as that of {\bf Fig. 1} but for $M_W$ = 80.22 GeV
               .

\end{description}
\end{document}